# Tuned bipolar oscillating gradients for mapping frequency dispersion of diffusion kurtosis in the human brain

Kevin B Borsos[* 1,2], Desmond HY Tse[2], Paul I Dubovan[1,2], Corey A Baron[1,2,3]

[1] Department of Medical Biophysics, Western University, London Ontario Canada
[2] Center for Functional and Metabolic Mapping, Western University, London Ontario Canada
[3] Imaging Laboratories, Robarts Research Institute, London Ontario Canada

* Corresponding author: Kevin B Borsos – kborsos@uwo.ca

## Abstract

**Purpose**: Oscillating gradient spin echo (OGSE) sequences have demonstrated an ability to probe time-dependent microstructural features, though they often suffer from low SNR due to increased echo times. In this work we introduce frequency tuned bipolar (FTB) gradients as a variation of oscillating gradients with reduced echo time and demonstrate their utility by mapping the frequency dispersion of kurtosis in human subjects.

**Methods**: An FTB oscillating gradient waveform is presented that provides encoding of 1.5 net oscillation periods thereby reducing the echo time of the acquisition. Simulations were performed to determine an optimal protocol based on SNR of kurtosis frequency dispersion - defined as the difference in kurtosis between pulsed and oscillating gradient acquisitions. Healthy human subjects were scanned at 7T using pulsed gradient and an optimized 23 Hz FTB protocol, which featured a maximum b-value of 2500 s/mm$^2$. In addition, to directly compare existing methods, measurements using traditional cosine OGSE were also acquired.

**Results**: FTB oscillating gradients demonstrated equivalent frequency dependent diffusion measurements compared to cosine modulated OGSE while enabling a significant reduction in echo time. Optimization and in vivo results suggest FTB gradients provide increased SNR of kurtosis dispersion maps compared to traditional cosine OGSE. The optimized FTB gradient protocol demonstrated consistent reductions in apparent kurtosis values and increased diffusivity in generated frequency dispersion maps.

**Conclusion**: This work presents an alternative to traditional cosine OGSE sequences enabling more time efficient acquisitions of frequency dependent diffusion quantities as demonstrated through in vivo kurtosis frequency dispersion maps.
2

# 1 Introduction

Diffusion MRI of the human brain is typically performed using pulsed gradient spin-echo (PGSE) sequences. While PGSE sequences enable efficient diffusion weighting, due to hardware constraints the range of accessible diffusion times remains limited to greater than ~30 ms. Complimentary to PGSE, oscillating gradient spin-echo (OGSE) diffusion gradients can achieve shorter effective diffusion times [1,2]. While the relationship between oscillation frequency $\omega$ and diffusion time is not well defined, it is generally accepted that shorter effective diffusion times can be achieved with increasing frequency compared to PGSE [3–5]. Consequently, OGSE has constituted a powerful tool in the investigation of time-dependent diffusion through monitoring of the ADC. Applications on this front include the probing of cell dimensions [6–11], surface-to-volume ratios [12,13] and additional microstructural characteristics such as packing [14,15], pore sizes [14] and extracellular space [16]. Recently, OGSE sequences utilized by Arbabi et al. [17] have also been used to investigate structural disorder in the human brain, confirming the predicted short-range disorder model [18].

However, despite the increased range of diffusion times and its ability to reveal otherwise inaccessible microstructural information, OGSE sequences typically suffer from an inherently lower signal-to-noise ratio (SNR) relative to PGSE [4]. The required time-modulation of the diffusion gradient results in multi-period waveforms that extend the TE of the acquisition. Resultant TEs are typically in excess of 110 ms, thereby severely reducing the SNR of DWIs and subsequently derived quantitative diffusion maps [19,20]. This makes high diffusion weighting applications such as quantifying the diffusion kurtosis [21,22] particularly challenging as the necessary b-values (>1500 s/mm$^2$) compound the attenuation of the diffusion signal. Moreover, the high b-values required for observing kurtosis can only be achieved by either utilizing lower oscillation frequencies or by increasing the number of oscillation periods (N) - both solutions require extending the diffusion gradient duration and thereby the TE. While recent advances in gradient hardware [23-25] can dramatically expand the range of b-values for a given frequency [26] (as the b-value scales with $1/\omega^3$), they do not significantly address the signal loss incurred due to extended TEs at the low frequencies ($\omega/2\pi < 50$ Hz) most often used for in vivo human imaging.

Hence there exists a need to further improve OGSE image quality and provide more accessible imaging parameters on a wider variety of systems. In this work we propose a novel frequency tuned bipolar (FTB) oscillating gradient waveform that reduces the TE of OGSE acquisitions without compromising the frequency probing characteristics of the technique. We present a direct comparison of this method to traditional OGSE and subsequently demonstrate its practicality by optimizing and acquiring kurtosis frequency dispersion maps of the human brain for the first time using a modern clinical gradient system.

## 2 Methods

2.1 Gradient Waveform Design

OGSE is conventionally performed with trapezoidal cosine modulated waveforms (Figure 1B) to maximize the achievable b-value [27]. Recently proposed by Hennel et al. [28], the conventional cosine sequence can be further modified by optimizing ramp times and reducing the spacing between the two diffusion gradients to the minimum allowable, effectively consolidating the two diffusion gradients into a single waveform. These changes were shown to produce more selective power spectra and increased diffusion weighting capabilities [28]. However this technique requires at least 2.5 net oscillation periods (N = 2.5 considering both sides of the refocusing RF pulse together) thus effectively trading a higher b-value for an increase in TE. Conversely, our proposed FTB waveform achieves shorter diffusion weighting durations by utilizing only N = 1.5 net oscillation periods over both sides of the refocusing RF pulse via two bipolar gradients that are tuned to achieve the desired frequency. This bipolar approach reduces the duration of the diffusion gradients and thereby significantly shortens the TE of the acquisition. A comparison between this new FTB implementation and typical N = 2 cosine modulated OGSE is presented in Figure 1, where similar spectral selectivity is demonstrated between the two methods (see Figures 1D and 1E).

The waveform is constructed by initially determining the duration of the second lobe (T) based on the target frequency $f$ (in Hz) according to the expression:

$$T = \frac{1}{2}\left(\frac{1}{f} - \tau_{RF} - 2\tau\right) \qquad (1)$$

where $\tau_{RF}$ is the shortest separation time required for the refocusing pulse and $\tau$ is the gradient rise time. This equation stems from an assumption that the central lobe of the three-lobe zeroth moment waveform (Figures 1A-C) will dominate the net frequency content of the diffusion weighting. Constraining the integral of the zeroth moment to be zero to eliminate any DC spectral components, the duration of the first lobe (L) can then be defined as (see Supporting Information for derivation):

$$L = \frac{1}{2}\left(-2T - \tau_{RF} - 7\tau + \sqrt{8T^2 + 8T\tau_{RF} + \tau_{RF}^2 + 32T\tau + 14\tau_{RF}\tau + 33\tau^2}\right) \quad (2)$$

The waveform's spectral fidelity can be compromised for large $\tau_{RF}$, which can result in deviations from the target frequency in addition to ringing and spectral broadening for frequencies higher than ~ 40 Hz (see Supporting Information Figure S2). Hence a narrow refocusing RF pulse is recommended to minimize the separation time.

2.2 Simulating SNR of Kurtosis Frequency Dispersion

Simulations of the diffusion signal were used to optimize the SNR of the kurtosis frequency dispersion ($\Delta K$); we define this quantity as:

$$\Delta K = K_{OGSE} - K_{PGSE} = K(\omega) - K(0) \quad (3)$$

The signal curves for both PGSE and OGSE acquisitions were generated using

$$S(b, TE, D, K) = S_0 e^{-bD + \frac{K b^2 D^2}{6}} e^{\frac{-TE}{T_2}} \quad (4)$$

at three different b-values: 0 s/mm², a maximum b-value that was varied, and a third intermediate b-value that was equal to half of the maximum. Due to considerations for clinical systems, the frequency was varied up to only 45 Hz. Gaussian noise was added to the calculated PGSE and OGSE signals (SNR = 35 at b = 0 s/mm²) upon which the magnitude of the noisy signals was then fitted with a non-negative least squares algorithm to recover ADC and kurtosis values. These values were then used to calculate $\Delta K$ according to equation 3. This procedure was repeated for 2000 iterations per each frequency/b-value combination. Subsequently the SNR of $\Delta K$ was estimated as the mean of these 2000 values divided by the standard deviation. The maximum b-values for OGSE and PGSE encoding were simulated for a gradient system with slew rate of 180 T/m/s and max gradient amplitude of 75 mT/m. To avoid the effects of higher order terms in the kurtosis signal expansion our simulations were limited to a maximum b-value of 2500 s/mm². The TE was chosen to be the minimum allowable for each frequency; accordingly, the TE is implicitly optimized through the optimization of the frequency.

Accurate simulation of the diffusion signal required *a priori* knowledge of the frequency dependence of the diffusivity D(ω) and kurtosis K(ω) to capture the time-dependence of both quantities. The model of the diffusion dispersion presented by Arbabi et al. [17] was used to infer D(ω) and the preliminary multi-frequency measurements of mean kurtosis presented by Yang et al. [30] were fitted to a power law model to obtain K(ω) = 0.93 – $0.0016\omega^{(0.78)}$ (see Supporting Information Figure S3).

Simulations were performed for FTB in addition to standard OGSE encoding using cosine N = 2 (Figure 1B) and N = 2.5 (similar to [28]) to observe the effects of adding additional periods. In addition to ΔK, the SNR of ΔD (defined similarly as ΔD = D(ω) - D(0) ) was also evaluated.

2.3 MRI Protocols

Three healthy participants were scanned using a head-only 7 Tesla MRI scanner (Siemens Magnetom 7T Plus, Erlangen Germany) capable of a maximum amplitude and slew rate of 80 mT/m and 400 T/m/s, respectively; however, to mitigate eddy currents the maximum gradient amplitude was limited to 75 mT/m and slew rate to 180 T/m/s. Approval for this study was granted by the Institutional Review Board at Western University and written informed consent was obtained from each participant prior to scanning.

According to the optimization results of Section 2.2, the optimized (highest SNR) kurtosis dispersion protocol consisted of PGSE and 23 Hz FTB oscillating gradients (L = 7.9 ms, T = 17.2 ms, $\tau_{RF}$ = 7.5 ms). The FTB and PGSE ($\Delta_{eff}$ = 36.1 ms) acquisitions were interleaved in a single scan using b-values of 0, 1250, 2500 s/mm$^2$ and implemented in vivo with a TE/TR of 91/6500 ms. In addition to human subjects, a multi-ADC diffusion phantom (CaliberMRI, Boulder Colorado, USA) was also scanned with this optimized protocol to validate dispersion measurements (Supporting Information Figure S4).

To compare to traditional cosine OGSE methods, participants were also scanned using N = 2 trapezoidal cosine modulated OGSE at the same frequency of 23 Hz. This required a longer minimum TE/TR of 128/7100 ms compared to the optimized FTB protocol; accordingly, an FTB sequence with TE/TR = 128/7100 ms was also acquired for comparison. For all cases, diffusion weighting was applied along 4-directions in a tetrahedral scheme [29] to enable maximum achievable b-values. The remaining imaging parameters were: FOV = 200×200 mm$^2$, matrix size = 100×100, 2174 Hz/Pixel bandwidth, 26 slices with no slice gap, 2 mm isotropic resolution, 8 averages, 6/8 partial Fourier EPI.

2.4 Image Processing and Analysis

Field dynamics were acquired using a field-monitoring system (Skope, Zurich Switzerland) and were integrated into an offline model-based image reconstruction algorithm to correct for eddy current distortions [31]. The complex images obtained from the reconstruction were phase-aligned by multiplying each image by $e^{-i\angle x}$, where x is the image after low pass filtering using a Kaiser window (width 40% of k-space). Principal component analysis based denoising was then applied to the phase-aligned complex images [32,33], and then the magnitudes of the images were extracted and processed with Gibbs ringing reduction using MRtrix3 [34].

DWIs from each shell were directionally averaged and fitted voxel-wise to the natural logarithm of the first exponential term of equation 3 with a non-negative least squares algorithm to extract ADC and kurtosis parameters. Prior to fitting, images were smoothed with a Gaussian filter using a full width at half maximum of 1 voxel. From the directionally averaged fitted data, spherical-mean ADC and kurtosis maps were generated for the PGSE and OGSE/FTB acquisitions separately. We note the mean apparent kurtosis, as used here, is distinct from the mean kurtosis which is derived from the diffusion kurtosis tensor [21]. Frequency dispersion maps of the ADC (ΔD) and kurtosis (ΔK) were then generated for each subject as the difference between PGSE and OGSE acquisitions as previously defined in Section 2.2.

To directly compare voxel values between acquisitions, cosine OGSE DWIs were registered to TE-matched FTB images by applying rigid body transformations using FSL [35,36]. Voxel-wise correlations of both ADC and kurtosis maps were used to assess the agreement between registered diffusion maps. To avoid the influence of implausibly high kurtosis values, only voxels with kurtosis < 3 were used. SNR maps for each subject were generated using the voxel-wise mean and standard deviation across the 8 b = 0 averages (prior to denoising). To compare SNR in white and gray matter, PGSE kurtosis values were used for segmentation of white and gray matter voxels based on the following thresholds: white matter: 1 < kurtosis < 1.5, gray matter: 0.4 < kurtosis < 0.8.

# 3 Results

## 3.1 Optimization Results

The optimal (highest SNR) protocol to acquire ΔK maps used 23 Hz FTB OGSE with a maximum b-value of 2500 s/mm$^2$ and TE of 91 ms (Figure 2A). The proposed FTB waveform can achieve 30% and 31% higher SNR for ΔK and ΔD maps respectively compared to traditional N ≥ 2 OGSE encodings (Figure 2B).

## 3.2 In vivo Results

TE matched FTB (FTB$_{TE}$) and cosine OGSE images exhibit qualitatively consistent ADC and kurtosis values (Figure 3A). Strong agreement between measured ADC values was observed when comparing the voxel-wise correlations across all subjects (Figure 3B). Similar agreement was also observed in apparent kurtosis values (Figure 3C), though more variation is observed due to the higher sensitivity of kurtosis to noise. Optimal FTB acquisitions (left side of Figure 3A) exhibited improved quality in ADC and kurtosis maps compared to the scans with longer TE. Cosine OGSE and FTB$_{TE}$ exhibited systematic overestimation of kurtosis compared to FTB, which is likely due to noise floor effects for the longer TE scans that had substantially lower SNR as seen in Figures 3A and 3D. Optimized (TE = 91 ms) FTB provided mean SNR improvements of 71% and 57% across subjects in white and gray matter respectively (denoted by the dashed and solid vertical lines in Figure 3D) when compared to OGSE and FTB$_{TE}$.

In generated frequency dispersion maps, decreasing kurtosis and increasing ADC is observed in FTB images compared to PGSE (Figure 4). In frontal white matter ROIs (see Figure 4C) mean differences between FTB and PGSE of approximately 11% and 15% are observed across subjects in apparent kurtosis and ADC values respectively. This frequency dependence was consistently observed over all subjects (Figure 5). Comparable image quality was observed across all subjects in both ΔK and ΔD maps (see Supporting Information Figure S5).

# 4 Discussion

## 4.1 Waveform Remarks

This work presents a frequency tuned bipolar OGSE gradient waveform that can significantly reduce the TE of OGSE acquisitions. By reducing the minimum number of oscillation periods to 1.5, significant gains in SNR were realized by achieving previously inaccessible TEs. Simulation results suggest this shorter FTB configuration to be ideal for measurements of diffusion ($\Delta D$) and kurtosis ($\Delta K$) frequency dispersion providing higher SNR for both types of parametric maps compared to OGSE performed with additional periods. This result agrees with previous findings from Arbabi et al. [17] that suggest waveforms with fewer periods provide higher SNR for $\Delta D$ maps. Moreover, the simulation results indicate the optimal frequency of 23 Hz was most influenced by differences in maximum b-value between frequencies. This effect is consistent with prior studies focusing on the optimization of kurtosis measurements [37,38] and suggests the SNR of $\Delta K$ maps will favor increasingly larger diffusion weighting.

We note however, that our optimization serves only to identify the most effective gradient scheme based on TE and b-value considerations. Further optimization will be required to determine the ideal number of shells/shell spacing as well as the number of diffusion directions. Notably, utilization of only 4 diffusion directions may introduce rotational variance [39], but this was required to maximize b-value in this preliminary work. However, the FTB waveform could be used with any number of directions, which would be feasible with stronger gradient systems.

The reduction in the number of OGSE periods facilitated by FTB may also prove useful by enabling the implementation of lower oscillation frequencies without drastically increasing the TE. For example, a 15 Hz cosine OGSE waveform would require a TE of 195 ms, but only 125 ms using FTB (see Supporting Information Figure S6). However, despite these advantages, the shortened duration of the FTB waveform further limits the range of possible b-values (Supporting Information Figure S6). As such, this method is likely best suited for lower oscillation frequencies ($f \leq 40$ Hz) where the reduction in b-value compared to traditional OGSE is negated by the increased b-value efficiency of a longer period waveform. Our technique will significantly benefit from the continued evolution of high-performance gradients whereby higher gradient amplitudes will offset the reduced b-value while still yielding shorter echo times. Additionally, FTB oscillating gradients can also be used in combination with recent advances in OGSE optimization such as the GRASE technique presented by Wu et al. [40] or spiral readouts [41] as demonstrated by Michael et al. [42] to further improve image and map quality.

4.2 In Vivo Findings

While $FTB_{TE}$ and OGSE exhibited highly correlated ADC and kurtosis measurements, limitations in this comparison included noise floor effects and the investigation of only a single *in vivo* frequency. However, preliminary Monte Carlo simulations in bundles of cylinders show extremely high correspondence of both ADC and kurtosis between the two methods for frequencies up to 60 Hz (Supporting Information Figure S7), further supporting equivalency of the methods. Nevertheless, the lack of experimental validation at multiple frequencies is a limitation of this work.

Experimentally, the ΔD maps demonstrate comparable diffusion dispersion to previously reported studies [17,5,42]. ΔK maps exhibit distinct contrast from ΔD, with larger magnitudes qualitatively observed in white matter (WM) relative to grey matter (GM) (Figures 4 and S5). This observation is consistent with the bimodal nature of the distributions in Figures 5C and 5D, and also with the only other documented preliminary measurements of kurtosis with OGSE in the human brain presented by Yang et al. [30]. The generally larger negative ΔK in WM suggests that, somewhat surprisingly, 23 Hz is a sufficiently high frequency to be in a regime where diffusion becomes increasingly Gaussian and kurtosis begins to vanish, similar to findings at short diffusion times in the ex vivo spinal cord and mouse brain [43,44]. Conversely, at even lower frequencies (i.e., longer diffusion times), it is expected that the trend of kurtosis with frequency will reverse due to the effects of permeability, contributing to both predicted [44,45] and observed [43,44,46,47] non-monotonic behavior. Given that the diffusion times accessible through OGSE are shorter than typical exchange times derived from the Kärger model, our measurements lie in the short diffusion time regime where exchange and permeability effects that act to reduce the apparent kurtosis likely do not have a significant role in these results [48,49].

While the biophysical interpretation of frequency dependent differences in kurtosis lies beyond the scope of this work, previous observations suggest ΔK may constitute a novel biomarker, enabling further insight into physiological conditions influencing the degrees of diffusion restriction - similar to the utility of ΔD [50-53]. Aggarwal et al. [44] observed significantly reduced ΔK in regions of local demyelination due to increased permeability, which suggests potential clinical uses for subtype identification or prognosis in demyelinating conditions like multiple sclerosis. Wu et al. [54] noted increases in ΔK corresponding to regions of severe edema in a mouse model of hypoxic ischemic injury, which suggests potential clinical applicability to stroke diagnosis or identification of the penumbra. That said, applications of ΔK in humans have yet to be explored due to the technical constraints that the FTB approach partially addresses. With the introduction of the FTB method and as remaining technical challenges continue to be addressed, the robust characterization of the relationship between kurtosis and frequency in humans and the application and interpretation of ΔK in patients will likely become possible.

## Conclusion

In this work a more time-efficient design for frequency-selective diffusion weighting is presented, enabling OGSE to be implemented with shorter echo times than previously possible. We demonstrate equivalency in diffusion contrast between existing OGSE methods and this new FTB approach and report preliminary frequency dispersion maps of the diffusion kurtosis in the human brain for the first time using a modern clinical gradient system. We anticipate as gradient hardware improves, the FTB framework presented here will become increasingly valuable as a method to further improve OGSE image quality through shorter echo times.

## Data Availability Statement

Anonymized participant data, simulation scripts and fitting routines used in this study are openly available for use and download from: *https://osf.io/d8pej/*

## Acknowledgments

This work was supported by the National Science and Engineering Research Council of Canada, the Canada First Research Excellence Fund to BrainsCAN and the Ontario Graduate Scholarship Program. The authors would like to thank Nico Arezza and Naila Rahman for helpful discussions regarding image registration.

# Bibliography

[1] Stepišnik J. Analysis of NMR self-diffusion measurements by a density matrix calculation. Physica B+ C 1981; 104:350–364.

[2] Callaghan PT, Stepišnik J. Frequency-domain analysis of spin motion using modulated-gradient NMR. J. Magn. Reson. 1995; 117:118–122.

[3] Does MD, Parsons EC, Gore JC. Oscillating gradient measurements of water diffusion in normal and globally ischemic rat brain. Magn Reson Med. 2003; 49:206–215.

[4] Xu J. Probing neural tissues at small scales: Recent progress of oscillating gradient spin echo (OGSE) neuroimaging in humans. J. Neurosci. Methods 2020; p. 109024.

[5] Baron CA, Beaulieu C. Oscillating gradient spin-echo (OGSE) diffusion tensor imaging of the human brain. Magn Reson Med. 2014; 72:726–736.

[6] Li H, Gore JC, Xu J. Fast and robust measurement of microstructural dimensions using temporal diffusion spectroscopy. J. Magn. Reson. 2014; 242:4–9.

[7] Xu J, Li H, Harkins KD, et al. Mapping mean axon diameter and axonal volume fraction by MRI using temporal diffusion spectroscopy. NeuroImage 2014; 103:10–19.

[8] Jiang X, Li H, Xie J, Zhao P, Gore JC, Xu J. Quantification of cell size using temporal diffusion spectroscopy. Magn Reson Med. 2016; 75:1076–1085.

[9] Xu J, Jiang X, Li H, et al. Magnetic resonance imaging of mean cell size in human breast tumors. Magn Reson Med. 2020; 83:2002–2014.

[10] Harkins KD, Beaulieu C, Xu J, Gore JC, Does MD. A simple estimate of axon size with diffusion MRI. NeuroImage 2021; 227:117619.

[11] Xu J, Jiang X, Devan SP, et al. MRI-cytometry: Mapping nonparametric cell size distributions using diffusion MRI. Magn Reson Med. 2021; 85:748–761.

[12] Reynaud O, Winters KV, Hoang DM, Wadghiri YZ, Novikov DS, Kim SG. Surface-to-volume ratio mapping of tumor microstructure using oscillating gradient diffusion weighted imaging. Magn Reson Med. 2016; 76:237–247.

[13] Novikov DS, Kiselev VG. Surface-to-volume ratio with oscillating gradients. J. Magn. Reson. 2011; 210:141–145.

[14] Gore JC, Xu J, Colvin DC, Yankeelov TE, Parsons EC, Does MD. Characterization of tissue structure at varying length scales using temporal diffusion spectroscopy. NMR Biomed. 2010; 23:745–756.

[15] Schachter M, Does M, Anderson A, Gore J. Measurements of restricted diffusion using an oscillating gradient spin-echo sequence. J. Magn. Reson. 2000; 147:232–237.

[16] Reynaud O, Winters KV, Hoang DM, Wadghiri YZ, Novikov DS, Kim SG. Pulsed and oscillating gradient MRI for assessment of cell size and extracellular space (POMACE) in mouse gliomas. NMR Biomed. 2016; 29:1350–1363.

[17] Arbabi A, Kai J, Khan AR, Baron CA. Diffusion dispersion imaging: Mapping oscillating gradient spin-echo frequency dependence in the human brain. Magn Reson Med. 2020; 83:2197–2208.

[18] Novikov DS, Jensen JH, Helpern JA, Fieremans E. Revealing mesoscopic structural universality with diffusion. PNAS 2014; 111:5088–5093.

[19] Dietrich O, Heiland S, Sartor K. Noise correction for the exact determination of apparent diffusion coefficients at low SNR. Magn Reson Med. 2001; 45:448–453.

[20] Saritas EU, Lee JH, Nishimura DG. SNR dependence of optimal parameters for apparent diffusion coefficient measurements. IEEE transactions on medical imaging 2010; 30:424–437.

[21] Jensen JH, Helpern JA, Ramani A, Lu H, Kaczynski K. Diffusional kurtosis imaging: the quantification of non-gaussian water diffusion by means of magnetic resonance imaging. Magn Reson Med. 2005; 53:1432–1440.

[22] Jensen JH, Helpern JA. MRI quantification of non-Gaussian water diffusion by kurtosis analysis. NMR Biomed. 2010; 23:698–710.

[23] Foo TK, Tan ET, Vermilyea ME, et al. Highly efficient head-only magnetic field insert gradient coil for achieving simultaneous high gradient amplitude and slew rate at 3.0 T (MAGNUS) for brain microstructure imaging. Magn Reson Med. 2020; 83:2356–2369

[24] Weiger M, Overweg J, Rösler MB, et al. A high-performance gradient insert for rapid and short-T2 imaging at full duty cycle. Magn Reson Med. 2018; 79:3256–3266.

[25] Huang SY, Witzel T, Keil B, et al. Connectome 2.0: Developing the next-

generation ultra-high gradient strength human MRI scanner for bridging studies of the micro-, meso-and macro-connectome. NeuroImage 2021; p. 118530.

[26] Tan ET, Shih RY, Mitra J, et al. Oscillating diffusion-encoding with a high gradient-amplitude and high slew-rate head-only gradient for human brain imaging. Magn Reson Med. 2020; 84:950–965.

[27] Van AT, Holdsworth SJ, Bammer R. In vivo investigation of restricted diffusion in the human brain with optimized oscillating diffusion gradient encoding. Magn Reson Med. 2014; 71:83–94.

[28] Hennel F, Michael ES, Pruessmann KP. Improved gradient waveforms for oscillating gradient spin-echo (OGSE) diffusion tensor imaging. NMR Biomed. 2021; 34:e4434.

[29] Conturo TE, McKinstry RC, Akbudak E, Robinson BH. Encoding of anisotropic diffusion with tetrahedral gradients: a general mathematical diffusion formalism and experimental results. Magn Reson Med. 1996; 35:399–412.

[30] Yang GK, Tan E, Fiveland E, Foo TK, McNab J. Measuring Time-Dependent Diffusion Kurtosis Using the MAGNUS High-Performance Head Gradient. In: Proceedings of ISMRM & SMRT Virtual Conference & Exhibition, 2020. p. 0962.

[31] Valsamis JJ, Dubovan PI, Baron CA. Characterization and correction of time-varying eddy currents for diffusion MRI. Magn Reson Med. 2022; 87:2209–2223.

[32] Does MD, Olesen JL, Harkins KD, et al. Evaluation of principal component analysis image denoising on multi-exponential MRI relaxometry. Magn Reson Med. 2019; 81:3503–3514.

[33] Veraart J, Novikov DS, Christiaens D, Ades-Aron B, Sijbers J, Fieremans E. Denoising of diffusion MRI using random matrix theory. NeuroImage 2016; 142:394–406.

[34] Tournier JD, Smith R, Raffelt D, et al. MRtrix3: A fast, flexible and open software framework for medical image processing and visualisation. Neuroimage 2019; 202:116137.

[35] Jenkinson M, Smith S. A global optimisation method for robust affine registration of brain images. Medical image analysis 2001; 5:143–156.

[36] Jenkinson M, Bannister P, Brady M, Smith S. Improved optimization for

the robust and accurate linear registration and motion correction of brain images. Neuroimage 2002; 17:825–841.

[37] Poot DH, Arnold J, Achten E, Verhoye M, Sijbers J. Optimal experimental design for diffusion kurtosis imaging. IEEE Trans. Med. Imaging 2010; 29:819–829.

[38] Gilani N, Malcolm PN, Johnson G. Parameter estimation error dependency on the acquisition protocol in diffusion kurtosis imaging. Appl. Magn. Reson. 2016; 47:1229–1238.

[39] Nilsson M, Szczepankiewicz F, Brabec J, et al. Tensor-valued diffusion MRI in under 3 minutes: an initial survey of microscopic anisotropy and tissue heterogeneity in intracranial tumors. Magn Reson Med. 2020; 83:608–620.

[40] Wu D, Liu D, Hsu YC, et al. Diffusion-prepared 3D gradient spin-echo sequence for improved oscillating gradient diffusion MRI. Magn Reson Med. 2021; 85:78–88.

[41] Lee Y, Wilm BJ, Brunner DO, et al. On the signal-to-noise ratio benefit of spiral acquisition in diffusion MRI. Magn Reson Med. 2021; 85:1924–1937.

[42] Michael ES, Hennel F, Pruessmann KP. Evaluating diffusion dispersion across an extended range of b-values and frequencies: Exploiting gap-filled OGSE shapes, strong gradients, and spiral readouts. Magn Reson Med. 2022; 00:1–14.

[43] Jespersen SN, Olesen JL, Hansen B, Shemesh N. Diffusion time dependence of microstructural parameters in fixed spinal cord. NeuroImage 2018; 182:329–342.

[44] Aggarwal M, Smith MD, Calabresi PA. Diffusion-time dependence of diffusional kurtosis in the mouse brain. Magn Reson Med. 2020; 84:1564–1578.

[45] Fieremans E, Novikov DS, Jensen JH, Helpern JA. Monte Carlo study of a two-compartment exchange model of diffusion. NMR Biomed. 2010; 23:711–724.

[46] Pyatigorskaya N, Le Bihan D, Reynaud O, Ciobanu L. Relationship between the diffusion time and the diffusion MRI signal observed at 17.2 tesla in the healthy rat brain cortex. Magn Reson Med. 2014; 72:492–500.

[47] Portnoy S, Flint J, Blackband S, Stanisz G. Oscillating and pulsed gradient diffusion magnetic resonance microscopy over an extended b-value range: implications for the characterization of tissue microstructure. Magn Reson Med. 2013; 69:1131–1145.

[48] Lee HH, Papaioannou A, Novikov DS, Fieremans E. In vivo observation and biophysical interpretation of time-dependent diffusion in human cortical gray matter. NeuroImage 2020; 222:117054.

[49] Nilsson M, Westen vD, Ståhlberg F, Sundgren PC, Lätt J. The role of tissue microstructure and water exchange in biophysical modelling of diffusion in white matter. Magn. Reson. Mater. Phys., Biol. Med. 2013; 26:345–370.

[50] Iima M, Yamamoto A, Kataoka M, et al. Time-dependent diffusion MRI to distinguish malignant from benign head and neck tumors. J. Magn. Reson. Imaging 2019; 50:88–95.

[51] Gao F, Shen X, Zhang H, et al. Feasibility of oscillating and pulsed gradient diffusion MRI to assess neonatal hypoxia-ischemia on clinical systems. J. Cereb. Blood Flow Metab. 2020; p. 0271678X20944353.

[52] Wu D, Zhang Y, Cheng B, Mori S, Reeves RH, Gao FJ. Time-dependent diffusion MRI probes cerebellar microstructural alterations in a mouse model of Down syndrome. Brain communications 2021; 3:fcab062.

[53] Maekawa T, Hori M, Murata K, et al. Differentiation of high-grade and low-grade intra-axial brain tumors by time-dependent diffusion MRI. Magnetic Resonance Imaging 2020; 72:34–41.

[54] Wu D, Li Q, Northington FJ, Zhang J. Oscillating gradient diffusion kurtosis imaging of normal and injured mouse brains. NMR Biomed. 2018; 31:e3917.

# Figures

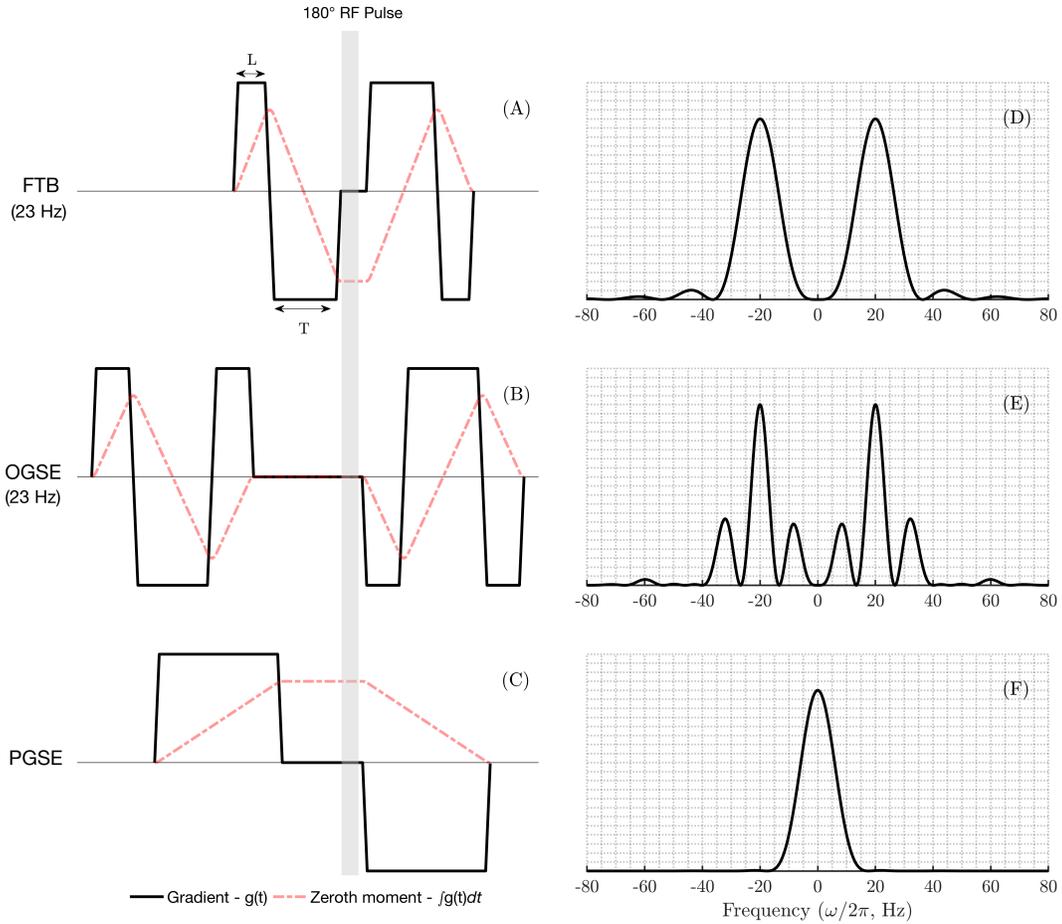

**Figure 1:** A 23 Hz frequency tuned bipolar (FTB) OGSE gradient waveform (A), conventional N = 2 cosine OGSE (B) and PGSE (C) gradient waveforms. Here $T$ and $L$ denote the flat top durations of the two lobes of the FTB gradient waveform, calculated from equations (1) and (2) respectively. Corresponding power spectra are shown in (D-F) demonstrating comparable spectral selectivity between FTB and N = 2 OGSE for the same target frequency. Also shown is the zeroth moment of each waveform (overlaid on A-C with dashed red lines). Uninterrupted periodicity of the zeroth moment is apparent in the FTB waveform that effectively has N = 1.5 periods (A) compared to N = 2 for OGSE (B) (visible from the zeroth moments). For OGSE, there is flexibility in the positioning of the waveform to the left of the refocusing RF pulse due to the extra time required for the EPI gradients on the right side; here, the position that maximizes spectral fidelity is used as described in [5]. Relative gradient amplitudes between FTB, OGSE, and PGSE are not to scale.

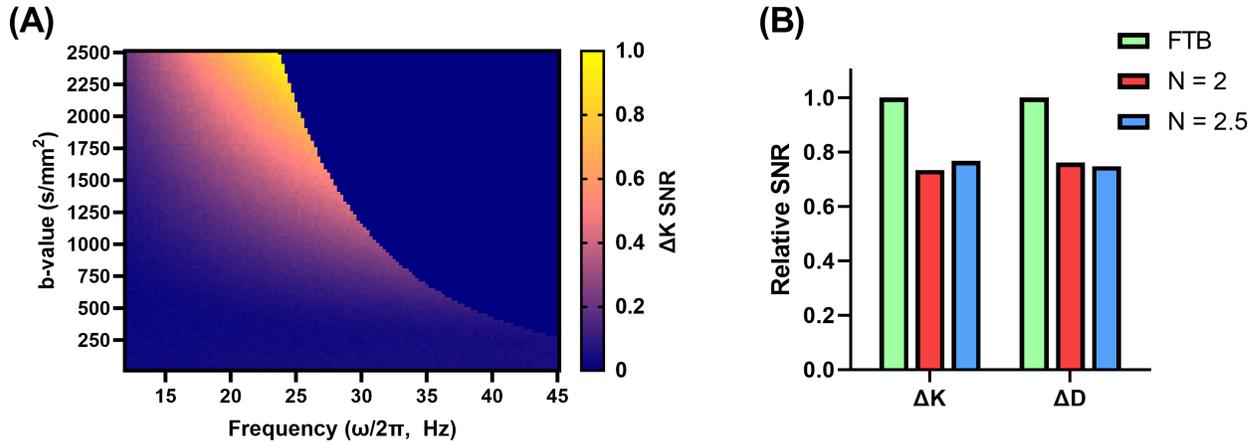

**Figure 2**: Simulation results for SNR of ΔK maps. (A) Heat map showing normalized simulated SNR of ΔK values for varying b-value and frequency combinations. Note the minimum TE was selected for each frequency, hence TE dependence is implicitly reflected through the frequency (TE varies from 125 ms for 15 Hz to 64 ms for 45 Hz). Maximum SNR occurs at a frequency of 23 Hz (TE = 91 ms) and b–value of 2500 s/mm$^2$. The masked region in the upper right denotes experimentally prohibited parameter combinations that would require a higher than maximum gradient amplitude. (B) Comparison of highest possible SNR of ΔK and ΔD maps achievable with FTB, N = 2 and N = 2.5 cosine OGSE gradient schemes. Values shown in (B) are normalized to the highest SNR achievable with FTB oscillating gradients for both ΔD and ΔK respectively and thereby denote relative differences (the absolute SNR of ΔD maps were approximately 4 times higher than ΔK).

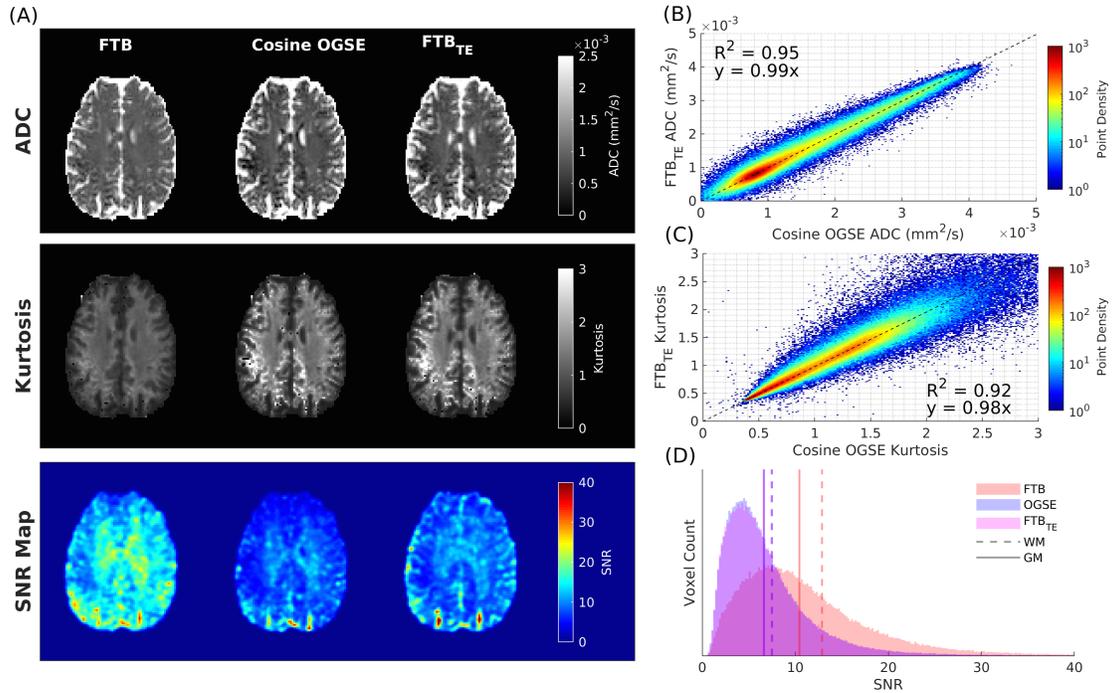

**Figure 3**: Comparing FTB to cosine OGSE. (A) comparison of ADC (top row) and apparent kurtosis maps (middle row) acquired using the following diffusion gradient schemes: FTB (TE = 91 ms), N = 2 cosine OGSE (TE = 128 ms) and FTB with the TE matched to the minimum possible with N = 2 cosine OGSE, $FTB_{TE}$ (TE = 128 ms). SNR maps (bottom row) were generated from 8 b = 0 images; for improved visualization these maps are smoothed with a Gaussian filter with a full-width at half-maximum of 3 voxels. (B,C) voxel-wise correlations of ADC (B) and kurtosis (C) values comparing measurements acquired using $FTB_{TE}$ and cosine OGSE across all imaging volumes from all subjects. Linear regressions are overlaid in black with equation and $R^2$ values shown on each plot demonstrating agreement of acquired diffusion metrics. As there are many data points, the logarithmic colormap reflects the local density of points per data marker. (D) Distribution of SNR values from all voxels across all subjects for FTB (red), TE matched FTB (magenta) and N = 2 cosine OGSE (blue); vertical lines denote the mean SNR in white (dashed) and gray (solid) matter for each scan (significant overlap is exhibited in the histograms and vertical lines for $FTB_{TE}$ and OGSE).

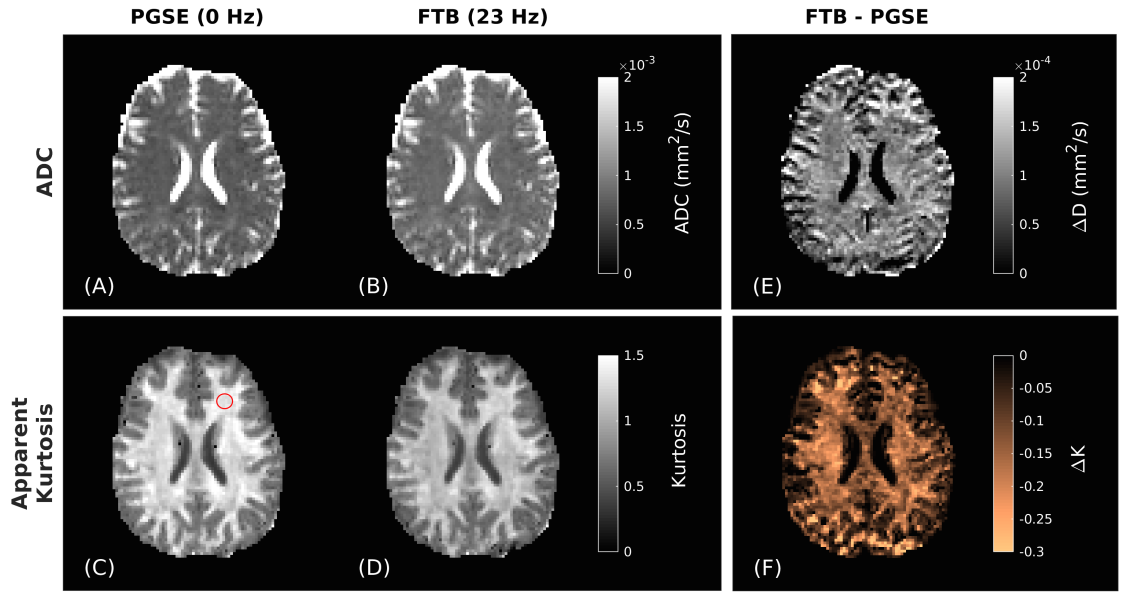

**Figure 4**: Generated ADC maps (A, B) and apparent kurtosis maps (C, D) from a healthy subject using both PGSE (A, C) and optimized FTB (B, D). An increase of ADC in parenchyma is observed in OGSE relative to PGSE while a reduction is demonstrated in apparent kurtosis. Also shown in (C) is an example region of interest used to calculate mean WM differences between OGSE and PGSE. Frequency dispersion maps of the ADC, ΔD (E) and kurtosis, ΔK (F) are also shown, demonstrating positive dispersion of the ADC and negative dispersion of the kurtosis.

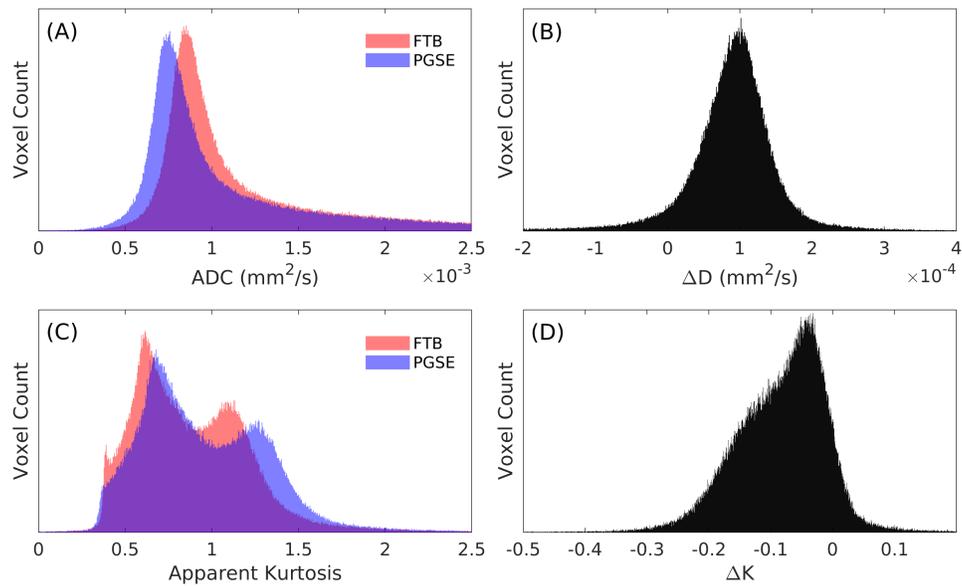

**Figure 5**: Histograms showing the distribution of ADC (A), apparent kurtosis (C) and frequency dispersion of ADC (B) and kurtosis values (D) of all voxels from all subjects. Elevated ADC values are observed in the FTB acquisitions relative to PGSE while kurtosis is reduced relative to PGSE; hence ΔD is observed to be predominately positive while ΔK is negative. The bimodal nature observed in the kurtosis (C) and ΔK (D) distributions is attributed to the presentation of voxels from both white and grey matter regions.

# List of Supporting Figure Captions

**Figure S1:** Schematic depicting the components of a typical FTB gradient waveform. Here τ denotes the gradient rise time (or ramp time), $\tau_{RF}$ the minimum width of the refocusing RF pulse (also referred to as separation time) and L and T the durations of the two lobes that comprise the gradient waveform.

**Figure S2:** Effect of increasing separation time ($\tau_{RF}$) on spectral fidelity of FTB oscillating gradient power spectra. Left column: power spectrum centroid frequency as a function of the target frequency for increasing separation times: 4 ms (top row), 8 ms (middle row) and 12 ms (bottom row). Significant deviations from the target frequency are observed as the separation time is increased – particularly at higher frequencies, though for typical RF pulse widths (5-8 ms) there is agreement between actual and target frequencies. Middle column: normalized example 60 Hz FTB gradient waveforms and corresponding zeroth moments (dashed red lines) for increasing separation times, 4 ms (top row), 8 ms (middle row) and 12 ms (bottom row). Interruption of the periodicity of the zeroth moment is apparent for increasing separation time. Right column: power spectra in arbitrary units (a.u.) of the 60 Hz FTB gradient waveforms shown in the middle column. Increased ringing is observed in spectra as separation time increases resulting in a higher centroid frequency than the set (target) frequency.

**Figure S3:** (A) Frequency dispersion models of the ADC, D(ω) and (B) kurtosis, K(ω) used for simulation of the diffusion signal in Section 2.2. The model for D(ω) was obtained from Arbabi et al. [17] while the model for K(ω) was generated from preliminary data presented in Yang et al. [30]. (C) The simulated maximum possible SNR of ADC and kurtosis (K) using FTB, N = 2 cosine OGSE and N = 2.5 OGSE; values are normalized to the highest SNR value (achieved with FTB for both ADC and K).

**Figure S4:** Mean ADC and kurtosis measurements obtained in a multi-ADC diffusion phantom (CaliberMRI, Boulder Colorado, USA) with both PGSE and frequency tuned bipolar OGSE (23 Hz) using a protocol identical to the in vivo scans. Equivalent ADC measurements are observed between OGSE and PGSE acquisitions (A) while zero kurtosis is also demonstrated for both acquisitions (B). Here error bars denote the standard deviation of the ROIs. Vial A and B denote two distinct concentrations of Polyvinylpyrrolidone solution designed to generate different ADC values; these values are specified by the manufacturer as 0.817 ×$10^{-3}$ and 1.109 ×$10^{-3}$ mm$^2$/s for vials A and B respectively.

**Figure S5**: Generated ΔD (A) and ΔK (B) maps from each subject. Comparable image quality is observed across participants though some B0 inhomogeneity induced distortions are apparent.

**Figure S6**: Comparison of maximum b-value (A) and minimum echo times (B) accessible using FTB and N = 2 Cosine OGSE. Values were calculated assuming a tetrahedral direction scheme on a system with maximum gradient amplitude and slew rate of 75 mT/m and 180 T/m/s respectively. FTB enables shorter echo times than cosine

OGSE at the expense of b-value, but the loss of b-value may not be relevant at low frequencies where the maximum b-value is still very large.

**Figure S7:** Monte Carlo simulations of diffusion in cylinders using Camino [Hall & Alexander, *IEEE Trans Med Imaging*, 2009; 28(9), 1354-1364] for OGSE frequencies from 12 Hz to 60 Hz. Three cases were simulated: OGSE using 5 periods with no gap for an RF pulse to approximate the ground truth (due to narrow bandwidth and small sidelobes), standard OGSE with 2 periods (one period on each side of the 180 degree RF pulse), and the proposed FTB method. (A) Example spectra for the 23 Hz case. (B) A distribution of cylinders was simulated that approximates the in vivo human brain, which leads to the shown contributions to MRI signal after accounting for the greater volumes of the larger cylinders [Baron et al. *Stroke*, 2015; *46*(8), 2136-2141, Aboitiz et al. *Brain research*, 1992; *598*(1-2), 143-153]. (C) Mean diffusivity and (D) kurtosis measured from the simulated diffusion MRI signals. Notably, despite differing spectra widths and sidelobe content, the measured frequency dependence of both diffusion parameters are nearly identical between all 3 cases across the range of frequencies.